\documentclass[12pt,a4paper,twoside]{article}
\usepackage{epsfig}
\usepackage{baltlat2}
\usepackage{wrapfig}
\usepackage{dcolumn}
\begin{document}
%%%%%%%%%%%

\titleb{OPTICAL SPECTROSCOPY OF RU\,CAM, A  PULSATING  CARBON STAR}
%Siin jarel peab tingimata olema tyhi rida, muidu laheb autori
%nimi pealkirjaga kokku!!!

\begin{authorl}
\authorb{T\~onu Kipper}{1} and
\authorb{Valentina G. Klochkova}{2} 
\end{authorl}

\begin{addressl}
\addressb{1}{Tartu Observatory, T\~oravere, 61602, Estonia; tk@aai.ee} 

\addressb{2}{Special Astrophysical Observatory RAS, Nizhnij Arkhyz, 369167,
             Russia; valenta@sao.ru} 
\end{addressl}

\submitb{Received ..., 2007}
\begin{abstract}
We analysed the high resolution spectra of a RU\,Cam, classified as W\,Vir
type star. The atmospheric parameters of RU\,Cam were estimated $T_{\rm
eff}$=5250\,K and $\log g$=1.0. The hydrogen deficiency of RU\,Cam was not
confirmed. The iron abundance, [Fe/H]=$-0.37$, is close to the solar one.
Abundances of most other elements are also close to normal. We found
considerable excesses of carbon and nitrogen: [C/Fe]=+0.98, [N/Fe]=+0.60. The
carbon to oxygen ratio is C/O$>$1. The carbon isotopic abundance ratio is
equal to $^{12}$C/$^{13}$C=4.5. For sodium a moderate overabundance
Na/Fe=+0.55 was obtained. For two moments of observations we found close
heliocentric velocity values, $v_{\rm r}$=$-21.7\pm0.8$ and
$-23.1\pm1.0$\,km\,s$^{-1}$. Both spectra contain a peculiar feature --
an emission component of Na\,I doublet which location agrees
with the radial velocity from the bulk of metallic lines. For our two observing
moments we found no dependence of radial velocities on the formation depth or
 on excitation energy
for metallic lines. 
\end{abstract}


\resthead{A pulsating star RU\,Cam}{T.\,Kipper, V.\,Klochkova }

{
\sectionb{1}{INTRODUCTION}
                                                                                
RU\,Cam is a variable star of W\,Vir type (Harris, 1985) with a
photometric period P\,$\approx 22.0^{\rm d}$ (Samus et al. 2004). Harris
(1985) estimated the star's distances from the Sun and from the galactic plane
as d=1.6 and z=0.7\,kpc. These distances are based on photometric data and well
agree with the value of the star's Hipparcos parallax $\pi$=0.59\,mas.
RU\,Cam has a rich history of photometric studies. Variability of its
radiaton was found a century ago by Ceraski (1907). After that a lot of
publications were devoted to studies of its peculiar photometric behavior
since the star has  variable pulsating magnitude and period. In
1965--1966 its irregular pulsation abruptly decreased in amplitude
from 1\,mag to about 0.1--0.2\,mag (Demers \& Fernie
1967) and  later exhibited a highly unstable and modulated light curve
(Kollath \& Szeidl 1993).

On the contrary to high photometric popularity, RU\,Cam has not been
 so popular for 
 spectroscopists. Among the first few was Sanford
(1928) who classified RU\,Cam as a carbon star. In the Catalogue of Carbon
Stars by Stephenson (1973) the star has a number CGCS\,6891.
The estimated spectral types of RU\,Cam are R0, K0var, C0.1, and
C3.2.e. Chemical
composition of RU\,Cam was studied first by Faraggiana
\& Hack (1967). Based on high resolution spectra and using the curve-of-growth
analysis, they concluded that the star's metallicity is close to normal. The
carbon excess of the star
was found to be  not larger than 2--3, the abundances of Ca, Ti, V, Ni,
and of rare
earths are also sligthly overabundant relative to Fe. Somewhat later
Wallerstein (1968) determined also the atmospheric parameters and 
reached similar
conclusions about the metallicity
 of RU\,Cam. These results were obtained with
photographic observations and with usage of  curve-of-growth analysis. In
order to use advantages of both modern spectroscopy and analytical
possibilities, we undertook a new research of RU\,Cam spectra.

\sectionb{2}{OBSERVATIONS}

Our high resolution spectra were taken with the Nasmyth Echelle
Spectrometer (Panchuk et al. 1999; Panchuk et al. 2002) of Russian 6\,m
telescope on Dec., 05 2006 (JD\,2454074.6) and on Feb.,07 2007
(JD\,2454138.5). The spectrograph was equipped with an image slicer
(Panchuk et al. 2003). As a detector a CCD camera with $2052\times2052$
pixels produced by the Copenhagen University Observatory was used. The
spectra for 2006 cover 516--669\,nm without caps until 610\,nm and the
spectra for 2007 
cover 452--602\,nm without gaps.

The spectra were reduced using the NOAO astronomical data analysis
facility IRAF. We describe the reduction procedure in Kipper \& Klochkova
(2005, 2006).

As measured from the Th-Ar comparison spectra the resolution is
$R\approx$\,42\,800 with FWHM of comparison lines about 7\,km\,s$^{-1}$.

\sectionb{3}{ANALYSIS and RESULTS}

\subsectionb{3.1}{Atmospheric parameters}

We mentioned above that RU\,Cam has been found to be a Pop.\,II Cepheid.
At the same time it is among the few which are carbon stars. If we assume
that RU\,Cam is a Pop.\,II cepheid one could derive from its light period
of 22 days $T_{\rm eff}$=5250\,K, $\log g$=1.2, $M_V=-2.4$, and
$M/M_{\odot}=0.6$ (Hall, 2000). These parameters were confirmed by Bergeat
et al. (2002), who found for RU\,Cam $T_{\rm eff}$=5215\,K, $M_{\rm
bol}$=$-1.8$ and pulsational mass $M_{\rm puls}$=0.57\,$M_{\odot}$. If we
adopt $M/M_{\odot}$=0.6 the surface gravity with found $M_{\rm bol}$ and
temperature will be $\log g$=1.44.

Kovtyukh et al. (1998) presented the calibrating relations between the
spectral line depths  and the excitation temperature for F--K
supergiants. Using their relations for 15 pairs of lines
we found $T_{\rm ex}$=5227$\pm$216\,K.
According to  these data the model (5250/1.5) from Kurucz's grid
(Kurucz, 1993) was chosen as a starting model. The independence of
abundances on excitation energy of the lines was confirmed for Fe\,I lines
(Fig.\,1, right panel). However, the ionization equilibrium of iron was not
satisfied
and the surface gravity was therefore reduced. As a result the 
final model (5250/1.0) was adopted. Also the microturbulent velocity
using Fe\,I lines was chosen $\xi_{\rm t}$=4.3\,km\,s$^{-1}$ (Fig.\,1,
left panel, illustrates a propriety of such a choise). Afterwards
the other elements showed slightly 
different $\xi_{\rm t}$ giving the error of $\xi_{\rm t}$
about 0.5\,km\,s$^{-1}$.

%%%%
\begin{figure}[h!]
\vskip2mm
\centerline{{\psfig{figure=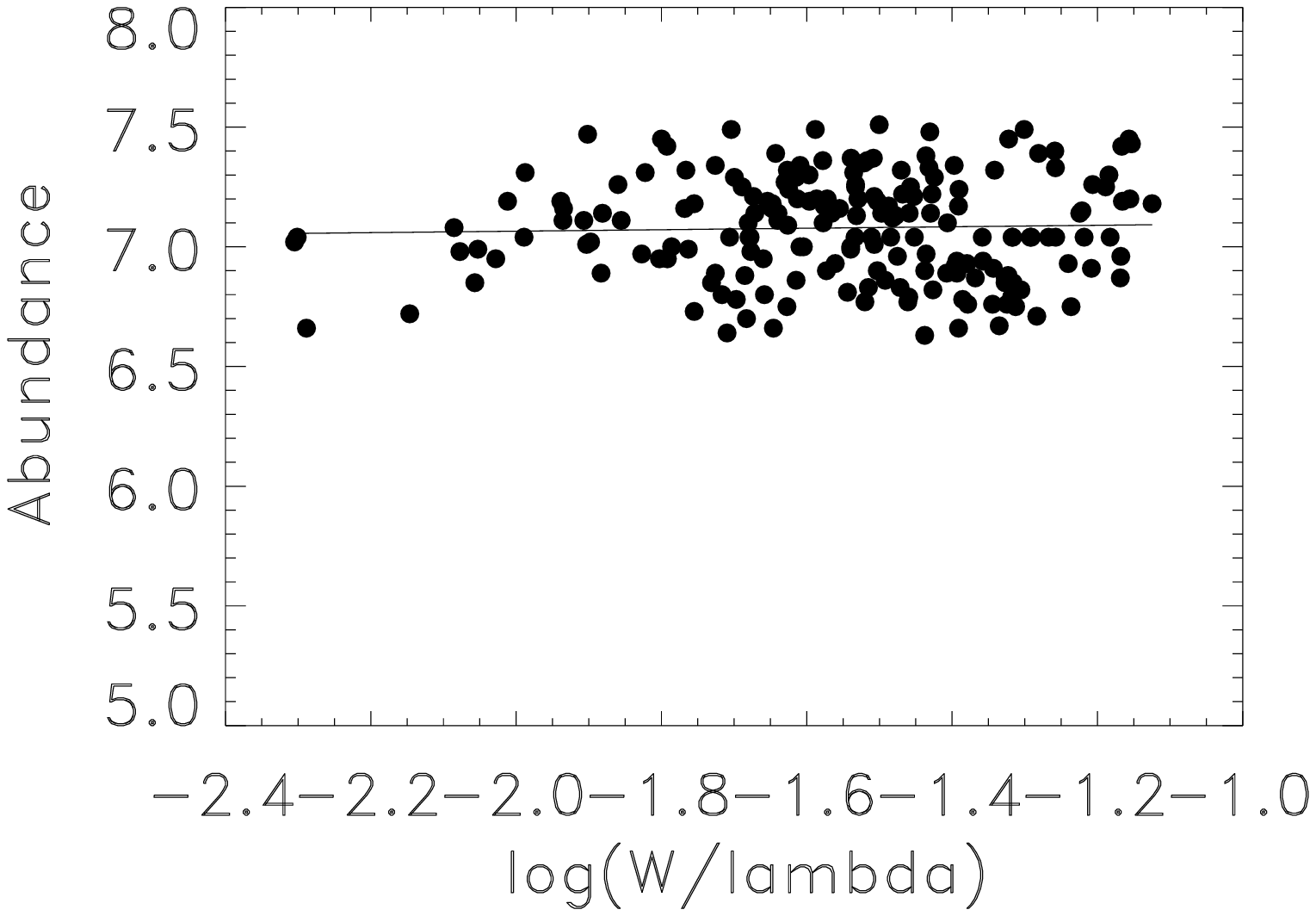,width=65truemm,angle=0,clip=}}
%%%%
{\psfig{figure=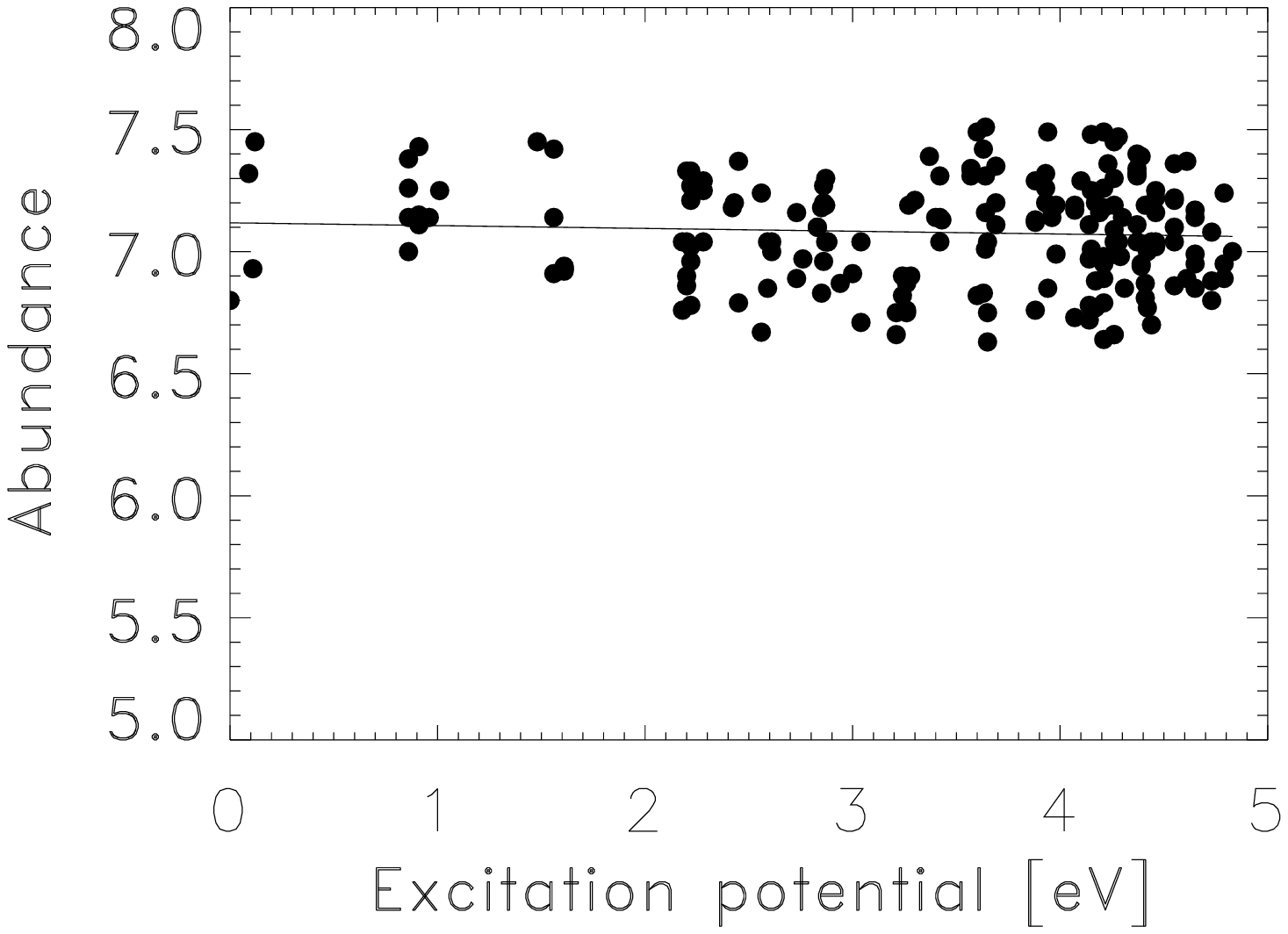,width=65truemm,angle=0,clip=}}}
\vskip2mm
\captionb{1}{ The dependence of the iron abundance for Fe\,I lines
on $W_\lambda/\lambda$ (left), and on lower excitation potential (right).
The microturbulent velocity is $\xi=4.3$\,km\,s$^{-1}$.}
\vskip2mm
\end{figure}

%%%%%%%%
%\vfill
%\eject
%%%%%%%

\subsectionb{3.2}{Chemical abundances}

The abundances were found using the Kurucz's program WIDTH5. Oscillator
strengts were taken from Thevenin (1989, 1990) except of those of C and
O, which were taken from Wiese et al. (1996).

\begin{center}
\vbox{\footnotesize
\begin{tabular}{lllrlr}
\multicolumn{6}{c}{\parbox{80mm}{\baselineskip=9pt
{\smallbf Table 1.}{\small\ The chemical composition of RU\,Cam. 
For comparison the abundances
in  V553\,Cen (Wallerstein \& Gonzalez, 1996) are given
	      in the last column.}}}\\[+4pt]
\tablerule
El. & $\log\varepsilon$ & $\log\varepsilon$ & [El/Fe]~ & Remarks & [El/Fe]  \\
 & Sun$^1$ & RU\,Cam & RU\,Cam &  & V553\,Cen  \\
\tablerule
C & 8.39 & $8.97\pm0.35$ & 0.98  & 8$^2$C\,I, C$_2$ bands & 0.88  \\
N & 7.78 & $8.00\pm0.30$ & 0.60  & CN bands               & 1.15  \\
O & 8.66 & $8.45\pm0.50$ & 0.16  & 1 O\,I, 2 [O\,I]       & 0.40   \\
Na& 6.17 & $6.35\pm0.17$ & 0.55  & 5 Na\,I                & 0.43  \\
Mg& 7.53 & $7.25\pm0.10$ & 0.09  & 6 Mg\,I                & 0.06  \\[5pt]
Si& 7.51 & $7.34\pm0.21$ & 0.20  & 16 Si\,I               & 0.17  \\
Ca& 6.31 & $5.94\pm0.31$ & 0.00  & 24 Ca\,I               & 0.20 \\
Sc& 3.05 & $2.71\pm0.21$ & 0.03  & 15 Sc\,II              & 0.14 \\
Ti& 4.90 & $4.58\pm0.25$ & 0.05  & 37 Ti\,I, 25 Ti\,II    & 0.05 \\
V & 4.00 & $3.82\pm0.23$ & 0.19  & 24 V\,I, 7 V\,II       & $-$0.20 \\[5pt]
Cr& 5.64 & $5.26\pm0.32$ &$-0.01$& 34 Cr\,I, 16 Cr\,II    & $-$0.18\\
Fe& 7.45 & $7.08\pm0.22$ &       & 191 Fe\,I,27 Fe\,II    & \\
Co& 4.92 & $4.79\pm0.28$ & 0.24  & 8 Co\,I                & 0.10\\
Ni& 6.23 & $5.84\pm0.26$ &$-0.02$& 52 Ni\,I               & -0.24 \\       
Cu& 4.21 & $4.08\pm0.06$ & 0.24  & 2 Cu\,I                & \\[5pt]
Zn& 4.60 & $4.02$        &$-0.21$& 1 Zn\,I                & \\
Y & 2.21 & $1.95\pm0.18$ & 0.11  & 7 Y\,II                & $-$0.22  \\
Zr& 2.59 & $2.27\pm0.30$ & 0.05  & 2 Zr\,I, 2 Zr\,II      &  \\
Ba& 2.17 & $1.72\pm0.08$ &$-0.08$& 4 Ba\,II               & 0.19 \\
La& 1.13 & $0.79\pm0.22$ & 0.03  & 4 La\,II               & $-$0.04 \\[5pt]
Ce& 1.58 & $1.03\pm0.33$ &$-0.18$& 9 Ce\,II               & $-$0.09\\
Pr& 0.71 & $0.33\pm0.07$ &$-0.01$& 3 Pr\,II               & \\
Nd& 1.45 & $1.12\pm0.16$ & 0.04  & 14 Nd\,II              & \\
Sm& 1.01 & $0.82$        & 0.18  & 1 Sm\,II               & \\    
\tablerule
\end{tabular}
}
\end{center}
\centerline{\parbox{110mm}{ \small\baselineskip=9pt
$^1$ Asplund et al. (2005), relative to $\log \varepsilon({\rm H})$,}}
\centerline{\parbox{110mm}{\small\baselineskip=9pt
$^2$ Number of used lines.}}
%%%%%%%%%%%%%%%%%%

\vskip2mm

As follows from Table\,1, the abundances determined for most chemical elements
are close to the chemical composition of the Sun (Asplund et al., 2005).
Metallicity of RU\,Cam is slightly decreased with [Fe/H]=$-0.37$. The result
 is
consistent with the  normal values of ratios [El/Fe] for a set of iron-group
elements: Sc, Ti, V, Cr, Ni. We note that the  metals with the high condensation
temperature (Ca and Sc) also have the  solar relative abundances: [Ca/Fe]=0.00
and [Sc/Fe]=0.03. This points  on ineffectiveness of selective depletion
 processes by dust formation. That agrees with nondetection  of dusty
IR--source associated with RU\,Cam. 

At the same time
we obtained the considerably modified content of CNO--triad: [C/Fe]= +0.98,
[N/Fe]=+0.60 and the carbon to oxygen ratio C/O$>1$. Overabundances of
carbon and nitrogen mean that matter
which had been  mixed into outer atmospheric
layers had been processed through He--burning. This points to an advanced
evolutionary stage of RU\,Cam. The obtained abundances indicate 
 that the matter
in RU\,Cam  suffered
helium burning followed by CN cycling and mixing to the surface of the
star. The carbon abundance found from the atomic carbon lines
was confirmed by synthesizing the C$_2$ Swan bands
$\log A({\rm C})=9.00\pm0.05$ and the nitrogen abundance 
was determined from the
CN red system bands $\log A({\rm N})=8.00\pm0.30$. The Bell's (1976)
line-list was used for spectrum synthesis. The carbon isotopic abundance ratio
$^{12}$C/$^{13}$C=4.5$\pm0.5$ was determined using the 
C$_2$ Swan system (1,0) bands
at 473.7 and 474.4\,nm
and (0,1) bands at 562.55 and 563.50\,nm.
 At these wavelengths Bell's list does not give
good wavelength match and therefore the list compiled by Alexander (1991)
was used. Earlier determinations of the carbon isotopic ratio for RU\,Cam were
by Climenhaga (1960), who found $^{12}$C/$^{13}$C=5.7 and Fraggania \&
Hack (1967) $^{12}$C/$^{13}$C=9.

In addition to CNO --excess, we obtained a moderate sodium overabundance,
[Na/Fe]=+0.55 which could be expalined by activity of Ne--Na cycle. One
could suspect that this Na-excess could be overestimated due to the
non-LTE effects in the atmosphere of cool supergiant. But, according to
Takeda et al. (2003), for the used lines the non-LTE effects are
practically insignificant being less than $-$0.10\,dex. For Na\,I D lines,
however, these effects are very large amounting to --1.0\,dex in the most
worse cases. This is one of the reasons why Na\,I D lines are not suitable
for abundance determinations. For the illustration these lines were
synthesized taking into account the hyperfine structure of the lines
(McWilliam et al. 1995). In Fig.\,3 the results are plotted. The used
abundance is by 0.9\,dex larger than found from weaker lines.

Abundances of all heavy metals (Y, Zr, Ba, La, Ce, Pr, Nd, Sm) are not
enhanced. Their relative contents are close to the solar ones. As a whole, the
chemical composition of RU\,Cam is not coincident with the chemical abundances
pattern typical for W\,Vir type stars in globular clusters.
Atmospheres of these evolved stars are metal-poor and enriched by helium,
carbon, and heavy metals of $s$--process (Gonzalez \& Wallerstein, 1994).
We have to note also that the chemical abundance pattern of RU\,Cam
differs from the chemical composition of W\,Vir itself -- the archetype of
population II cepheids. According to Barker et al. (1971) W\,Vir is a
metal-poor star, its metallicity [Fe/H]=$-1.1$, its content of heavy metals
 is essentially decreased relative to metallicity: [Met/Fe]=$-$2.2.

Lloyd Evans (1983) selected a small group (7 stars) of pulsating
 stars having a
carbon excess and called them carbon cepheids. These stars show strong
absorbtion bands of C$_2$, CH, CN and the absence of any enhancement of
the heavy metals produced in $s$--process. 
 RU\,Cam was listed among these objects.
Two  of this group members V553\,Cen and RT\,TrA were
 studied by Wallerstein \& Gonzalez (1996) and Wallerstein et al. (2000).
In order to  compare the results
 we present in the last column of Table\,1  the chemical abundances
of V553\,Cen obtained by these
authors. The results for RT\,TrA are very close to that of V553\,Cen. The
comparison shows almost coinciding abundance pattern. The carbon isotopic
abundance ratio for V553\,Cen and RT\,TrA $^{12}$C/$^{13}$C=4.0--5.0 is
also close to that of RU\,Cam. Both these stars have been associated
with shorter period subclass of Pop.\,II cepheids BL\,Her. They are somewhat
hotter and less luminous than RU\,Cam. At the moment no evolutionary
sequences could predict their chemical composition.

\sectionb{3.3}{Hydrogen deficiency}

Bergeat et al. (2002) when deriving the parameters of RU\,Cam added a HdC
label to it. However we found that the H$\alpha$ and H$\beta$ lines are quite
normal for  early K spectral type. We compare the profiles of the H$\alpha$ 
and H$\beta$ lines in RU\,Cam with those 
 in Arcturus (K1.5III) spectrum (Fig.\,2).
On our request RU\,Cam was observed by K.\,Annuk on June, 06 2007 with
the Tartu Observatory 60'' telescope with resolution $R\approx2500$ near
the CH G-band near 430\,nm. Comparison of this spectrum with the synthesized CH
band spectrum  gave satisfactory fit. For synthesizing we used the Bell's
line-list, model (5250/1.0) and normal H content. We therefore judge that
RU\,Cam is not hydrogen deficient.
 A similar result was obtained
by Faraggiana \& Hack (1967). 
%%%%%%%%%%%%%%%%%%%%%%%%%%%%%

\begin{figure}[hbtp]
\vskip2mm
\centerline{{\psfig{figure=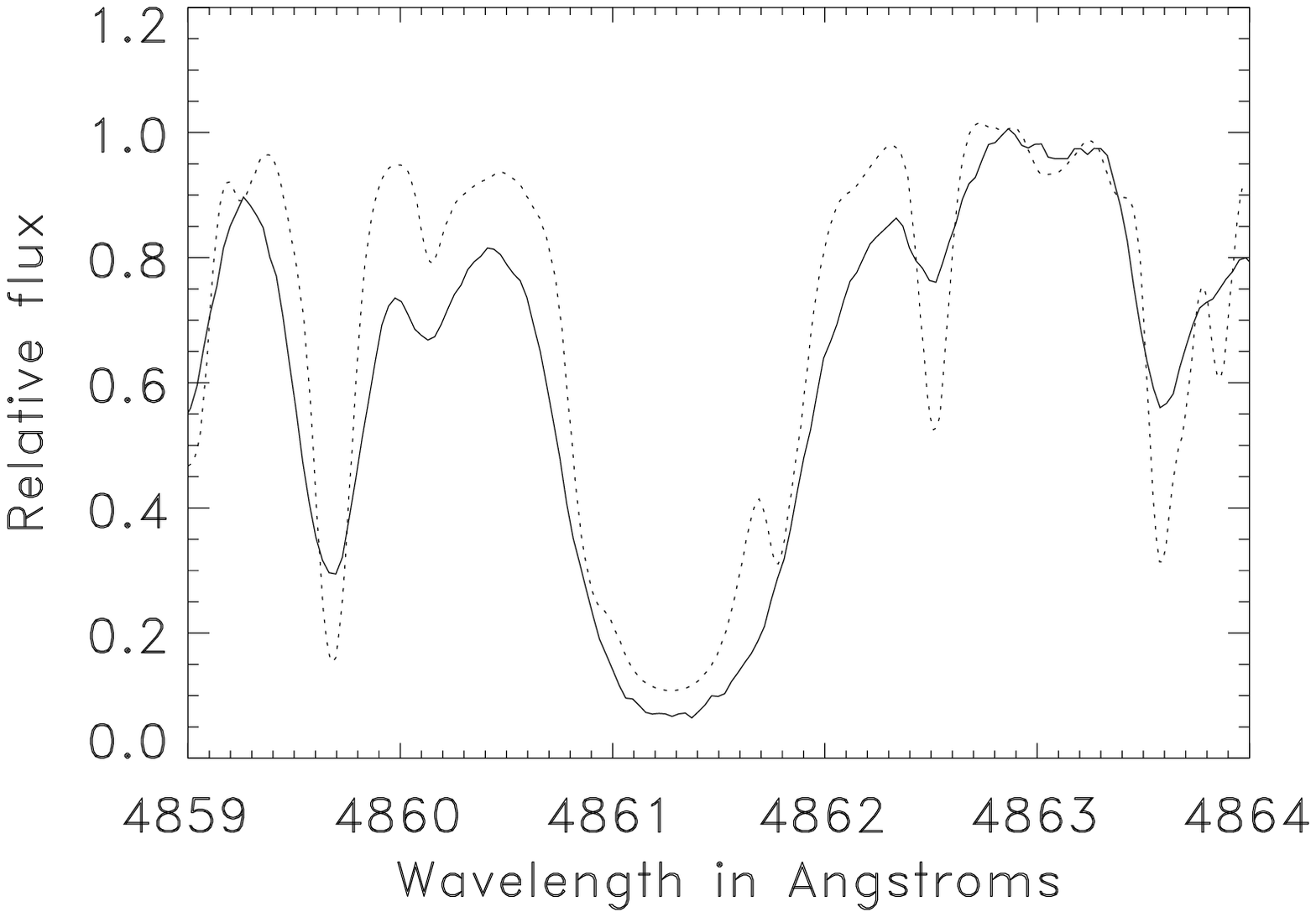,width=65truemm,angle=0,clip=}}
%%%%
{\psfig{figure=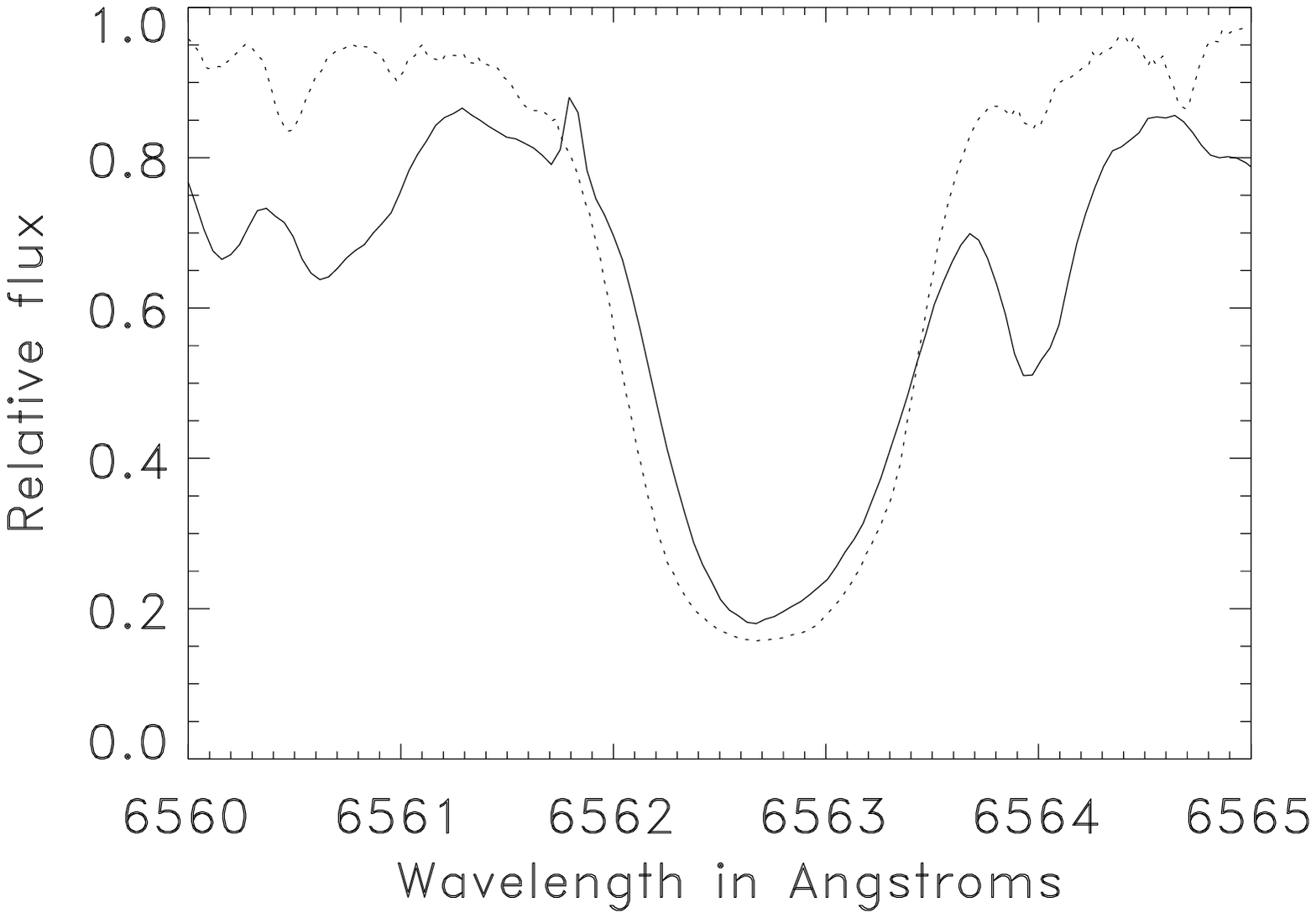,width=65truemm,angle=0,clip=}}}
\vskip2mm
\captionb{2}{The comparison of the H$\beta$ and H$\alpha$ lines in
the spectra of RU\,Cam (solid line)  and Arcturus.}
\vskip2mm
\end{figure}

\sectionb{4}{RADIAL VELOCITIES}

In the following we call the spectra obtained on Dec., 05 2006 for shortness as
SpI, and the ones obtained on Feb., 07 2007 as SpII. The corrections due to
the solar motion we adopted for SpI is $+9.37$\,km\,s$^{-1}$ and for SpII
$-$12.64\,km\,s$^{-1}$. To improve the accuracy we selected only weakly
blended lines. For SpI we measured 158 lines and for SpII -- 150 lines.
All results concerning radial velocities are presented in Table\,2. The
average radial velocity obtained from metallic lines for SpI was
$v_{\rm r}$=$-21.7\pm0.8$\,km\,s$^{-1}$ and
$v_{\rm r}$=$-23.1\pm1.0$\,km\,s$^{-1}$ for SpII. The errors indicated
 are the weighted by the number of lines
standard deviations of measurements for different metals. Coincidence of
our two $v_{\rm r}$ values is explained by similar phase of our
observations since the observing moments are separated by
approximately 64 days what is close to 3P.

Our $v_{\rm r}$ values agree very well with the earlier published data.
Wallerstein \& Crampton (1967) found after RU\,Cam has ceased its light
variations $v_{\rm r}=-22.9\pm 2.2$\,km\,s$^{-1}$. Much earlier Sanford had
(1928) found the velocity variation $v_{\rm r} \approx -3\div -37$ and the
systemic velocity --23.9\,km\,s$^{-1}$. Later Barnes et al. (1988)
confirmed this interval of $v_{\rm r}$ variability.

%%%%%%%%%%%%%%%%%%%%%%%%%%%%%%%%%%%%%
 If the radial velocity is
completely caused by Galactic rotation at the distance of 1.7\,kpc in
direction of RU\,Cam it would be --14\,km/s. Our measured velocity
is reached at the distance of 2.7\,kpc. This is also the upper limit
posed by Wallerstein (1968). If the distance is  so large the star's
bolometric magnitude would be $M_b=-3.9$. This means that $\log g$ should
be lowered  0.5 dex, which is not impossible considering our 
errors.
 This luminosity is already higher than
the carbon stars formation limit by third dredgeup on TP-AGB. But
this is not certainly true for V553\,Cen and RT\,TrA. 
It is doubtful if the shorter and longer period
carbon cepheids have different origins and exactly the same chemistry.

The Na\,I\,D lines in the RU\,Cam spectra have  composite profiles.
First, they show a two-components structure (Fig.\,3). We propose that there
are emission components close to the core of both D--lines. As follows from
Table\,2, location of Na\,I--emission to $v_{\rm r}$ practically
coincides with the average $v_{\rm r}$ value from numerous metallic
lines. Secondly,  the red wings of Na\,I profiles are
slightly sharper than the blue ones. Earlier Faraggiana \& Hack (1967)
detected emission core in the H and K lines of Ca\,II. These authors
believed that emissions are of chromospheric origin. Indeed, due to the star's
location at fairly high galactic latitude, $l$=+29$^o$, its interstellar
extinction does not exceed 0.01$^m$ (Wamsteker, 1966). This means that we
see Na\,I lines without interstellar components. This suggestion is
confirmed by the 
comparison of the observed and theoretical spectra near the
Na\,I doublet (Fig.\,3). It is natural to suggest for a pulsating
star that emission components of D--lines Na\,I indicate the presence of a
shock wave in the stars atmosphere. But permanency of both intensity and
location of emissions is doubtful in the framework
of this proposal. In accordance with 
Faraggiana \& Hack (1967), we are apt to think that a gaseous envelope
reveals itself in the Na\,I emissions.

\begin{figure}[h!]
\vskip2mm
\centerline{{\psfig{figure=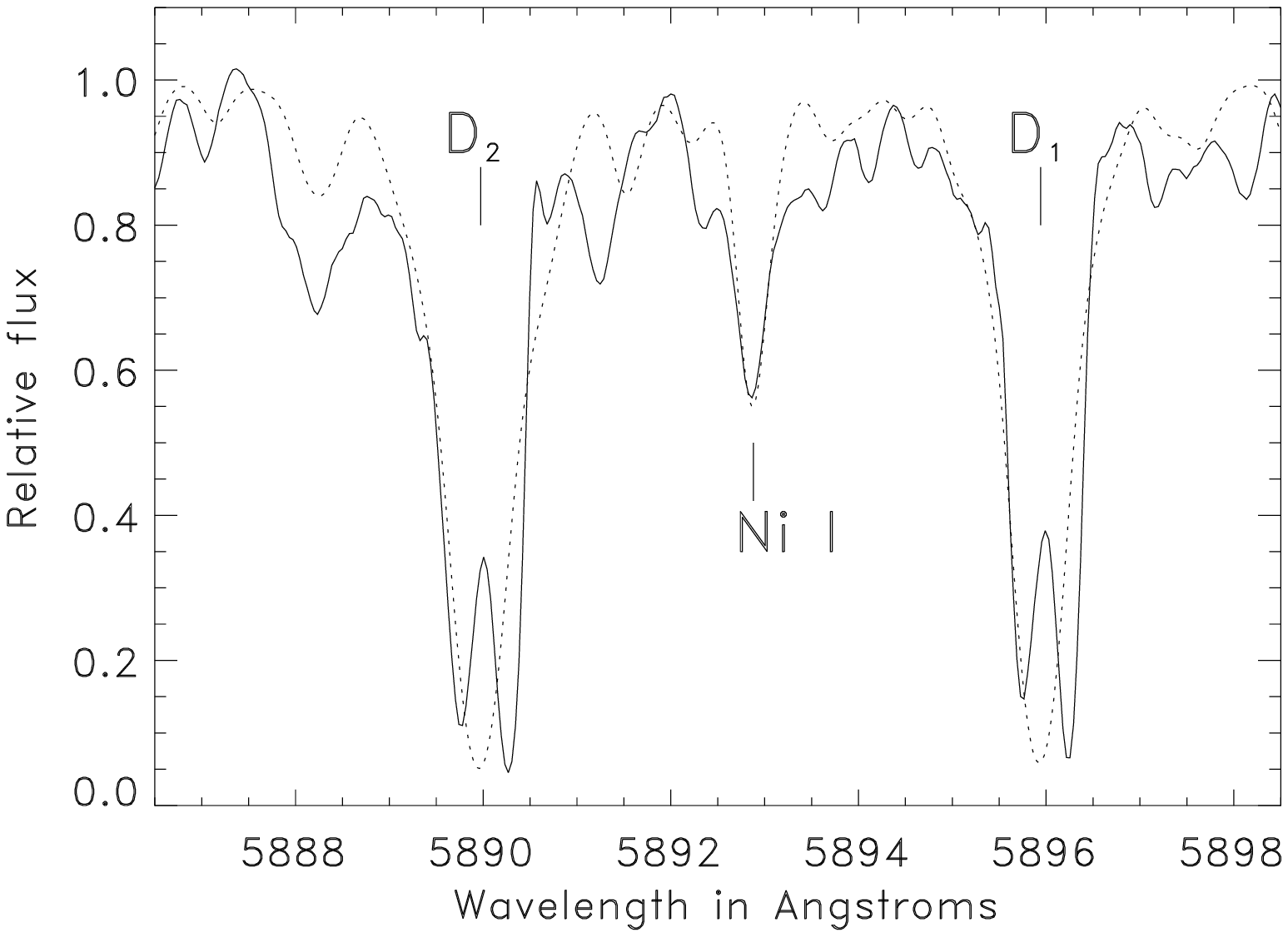,width=110truemm,angle=0,clip=}}}
%%%%
\captionb{3}{ The region of RU\,Cam spectrum near the Na\,I doublet (solid
              line) and LTE synthetic spectrum with the model (5250/1.0)
	      and the sodium abundance $\log A({\rm Na})=7.25$. The hyperfine
	      splitting of Na\,I lines is taken into account. }
\vskip2mm
\end{figure}

\begin{center}
\vbox{\footnotesize
\begin{tabular}{lccccc}
\multicolumn{6}{c}{\parbox{100mm}{\baselineskip=9pt
{\smallbf Table 2.}{\small\ Radial velocities of RU\,Cam
  for two observing dates derived from 
                   various spectral features.
                   For metallic
		   lines the number of measured lines is indicated
		   in parentheses.
 }}}\\ [+4pt]
\tablerule
{\raisebox{-2.0ex}{Date}}&&&$v_{\rm r}$\,[km\,s$^{-1}$]&& \\
         &Metals&H$\alpha$&H$\beta$&D$_2$(em)&D$_1$(em) \\    
\tablerule
Dec., 05 2006 & $-21.7\pm0.8$(158) & $-20.6$&       &$-19.3$&$-18.1$ \\
Feb., 07 2007 & $-23.1\pm1.0$(150) &        &$-19.4$&$-19.4$&$-18.7$ \\
\tablerule
\end{tabular}
}
\end{center}

Faraggiana \& Hack (1967) found an evidence of $v_r$ stratification:
spectral lines of different excitation give different $v_r$. Having
numerous accurate $v_r$ values, we check whether the radial velocities
depend on excitation energy or the formation depth of the lines. For this
aim we used Fe\,I, Fe\,II, Ti\,I and Ti\,II lines. As a result, for two
moments of observations we found no dependence on formation depth as
defined in Kurucz's program WIDTH5 or on excitation energy for these
lines. 
As is evident in Fig.\,2 the H$\alpha$ profile is asymmetrical and if the blue
wing is mirrored, then $v_{\rm r}$=$-$24.9\,km\,s$^{-1}$. 
If the full line is fitted  $v_{\rm r}$=$-$20.6\,km\,s$^{-1}$.
This asymmetry could well be caused by the emission in red wing.
The H$\beta$ line is more
or less symmetrical and $v_{\rm r}$=$-$19.4\,km\,s$^{-1}$.
The metal lines are more blueshifted than the hydrogen lines, but taking into
account the accuracy of $v_{\rm r}$ measurements, we may conclude that the
position of both H\,I lines is consistent with the value of $v_{\rm r}$
derived from numerous metallic lines.

\sectionb{5}{CONCLUSIONS}

Based on the high resolution spectra of a carbon star RU\,Cam, we obtained
its atmospheric parameters $T_{\rm eff}$=5250\,K, $\log g$=1.0,
$\xi_{\rm t}=4.3\pm0.5$\,km\,s$^{-1}$,  and
detailed chemical composition. As a result, the hydrogen deficiency of
RU\,Cam was not confirmed. The iron abundance, [Fe/H]=$-0.37$, is close to
the solar one. The abundances of most other elements are also
 close to normal. We
obtained considerably altered abundances of carbon and nitrogen:
[C/Fe]=+0.98, [N/Fe]=+0.60. The carbon to oxygen ratio is C/O$>1$. The sodium
overabundance, Na/Fe=+0.55, is real since the  non-LTE effects for the studied
Na\,I lines  are small. As a whole the chemical composition of RU\,Cam is
not coincident with the  chemical abundances pattern typical for
W\,Vir type stars. 
% Earlier,
%Wannier et al. (1990) had noted that  RU\,Cam is  a star with unusual
%evolutionary state. 

The heliocentric velocity values
$v_{\rm r}$= $-21.8\pm1.8$ and $-23.2$\,km\,s$^{-1}$ taken for 2
close photometric phases are coincident within the error box.

Both spectra of RU\, Cam contain a peculiar feature -- an emission
component  of Na\,I doublet whose location agrees with the radial
velocity from the bulk of metallic lines.

As a whole, taking 
into account the position above galactic plane, the close to the solar
metallicity of RU\,Cam, details of chemical composition and value of its 
systemic velocity $v_{\rm r}$=$-$24\,$^{-1}$, we may conclude that this
far evolved star belongs to thick disc population of Galaxy.

\vskip5mm

ACKNOWLEDGEMENTS.\ This research was supported by the Estonian Science
Foundation grant nr.~6810 (T.K.). V.G.K. acknowledges the support from the
programs of Russian Academy of Sciences ``Observational manifestations of
evolution of chemical composition of stars and Galaxy'' and ``Extended
objects in Universe''. V.G.K. also acknowledges the support by Award
No.\,RUP1--2687--NA--05 of the U.S. Civilian Research \& Development
Foundation (CRDF).

\goodbreak

\References{}

\refb
Alexander D.R. 1991, private communication
\refb
Asplund M., Grevesse N., Sauval J. 2005, ASP Conf. Ser., 336, 25
\refb
Barker T., Baumgart L.D., Butler D., et al. 1971, ApJ, 165, 67
\refb
Barnes III T.G.,  Moffett T.J., Slovak M.H. 1988, ApJ,  66, 43
\refb
Bell R.A. 1976, private communication,
 http://ccp7.dur.ac.uk/ccp7/DATA/lines.bell.tar.Z
\refb
Bergeat J., Knapik A., Rutily B. A\&A, 2002 390, 967
\refb
Ceraski W. Astron. Nachr.  1907, 174, 79
\refb 
Climenhaga J.L. Publ. DAO, 1960, 11, 307
\refb
Demers S., Fernie J.D. ApJ, 1966, 144, 440
\refb
Faraggiana R., Hack M. Zeitshrift f\"ur Astrophysics, 1967, 66, 343
\refb
Gonzalez G.\& Wallerstein G. AJ, 1994, 108, 1325
\refb
Hall D.A. In: Allen's Astrophysical Quantities, ed. A.N.\,Cox, Springer, 2000, P.400
\refb
Harris H.C. 1985, AJ, 90, 756
\refb
Kipper T., Klochkova V.G. 2005, Baltic Astronomy, 14, 215
\refb
Kipper T., Klochkova V.G. 2006, Baltic Astronomy, 15, 531
\refb
Kollath Z., Szeidl B. 1993, A\&A, 277, 62
\refb
Kovtyukh V.V., Gorlova N.I., Klochkova V.G. Astron. Letters, 1998, 24, 372
\refb
Kurucz R.L., 1993, SAO, Cambridge, Kurucz CDROM18
\refb
Lloyd Evans T. Observatory, 1983, 103, 276
\refb
McWilliam A., Preston G.W., Sneden C., Searle L. 1995, AJ 109, 2757
\refb
Panchuk V.E., Klochkova V.G., Naidenov I.D. 1999, Preprint
Spec. AO, No 135
\refb
Panchuk V.E., Piskunov N.E., Klochkova V.G., et al. 2002, Preprint
Spec. AO, No 169
\refb
Panchuk V.E., Yushkin M.V., Najdenov I.D. 2003, Preprint Spec. AO, No 179
\refb
Samus N.N., Durlevich O.V., et al. 2004, Combined Catalog of Variable
          Stars (GCVS4.2, 2004 Ed.)
\refb
Sanford R.F. ApJ,  1928, 68, 408 
\refb
Stephenson C.B.  General  catalogue  of  galactic carbon  stars. Publ.
            Warner \& Swasey Obs., 1973, 1, 1  
%\refb
%Torres--Peimbert S., Wallerstein G. 1966, ApJ, 146, 724
\refb
Takeda Y., Zhao G., Takada-Hidai M., Chen Y.-Q., Saito Y., Zhang H.-W. 2003,
      Chin. J Astron. Astrophys. 3, 316.
\refb
Thevenin F. 1989, A\&AS, 77, 137
\refb
Thevenin F. 1990, A\&AS, 82, 179
\refb
Wallerstein G. ApJ, 1968, 151, 1011
\refb
Wallerstein G., \& Crampton D. ApJ, 1967, 149, 225
\refb
Wallerstein G., \& Gonzalez G.  MNRAS, 1996, 282, 1236
\refb
Wallerstein G., Matt S., Gonzalez G. MNRAS, 2000, 311, 412
\refb
Wamsteker W., 1966, IBVS, No.\,128
%\refb
%Wannier P.G., Sahai R., Andersson B.-G., Johnson H.R. ApJ, 1990, 358, 251
\refb
Wiese W.L., Fuhr J.R., Deters T.M. 1996, J. Chem. Ref. Data, Mono. 7

\end{document}